\documentclass[]{interact}
\usepackage{orcidlink}
\usepackage{graphicx}% Include figure files
\usepackage{dcolumn}% Align table columns on decimal point
\usepackage{bm}% bold math %\bibliographystyle{apsrev}
\usepackage{bbold}
\usepackage{CJK}
\usepackage{amsthm,amssymb,bbm}
\usepackage{epstopdf}% To incorporate .eps illustrations using PDFLaTeX, etc.
\usepackage[caption=false]{subfig}% Support for small, `sub' figures and tables
\newcommand{\ra}{\rangle}
\newcommand{\la}{\langle}
\usepackage[numbers,sort&compress]{natbib}% Citation support using natbib.sty
\bibpunct[, ]{[}{]}{,}{n}{,}{,}% Citation support using natbib.sty
\renewcommand\bibfont{\fontsize{10}{12}\selectfont}% Bibliography support using natbib.sty
\makeatletter% @ becomes a letter
\def\NAT@def@citea{\def\@citea{\NAT@separator}}% Suppress spaces between citations using natbib.sty
\makeatother% @ becomes a symbol again

\theoremstyle{plain}% Theorem-like structures provided by amsthm.sty

\theoremstyle{definition}

\theoremstyle{remark}

\begin{document}

\title{Managing the Three-Party Entanglement Challenge}

\author{
\name{Songbo Xie and J.H. Eberly \thanks{Email: sxie9@ur.rochester.edu}}
\affil{Center for Coherence and Quantum Science, and Department of Physics and Astronomy, University of Rochester, Rochester NY 14627 USA.}
}

\maketitle

\begin{abstract}
We introduce the challenges of multi-party quantum entanglement and explain a recent success in learning to take its measure. Given the widely accepted reputation of entanglement as a counter-intuitive feature of quantum theory, we first describe pure-state entanglement itself. We restrict attention to multi-party qubit states. Then we introduce the features that have made it challenging for several decades to extend an entanglement measure beyond the 2-qubit case of Bell states. We finish with a description of the current understanding that solves the 3-qubit entanglement challenge. This necessarily takes into account the fundamental division of the 3-qubit state space into two completely independent sectors identified with the so-called GHZ and $W$ states. 
\end{abstract}


\section{Introductory Background}
Entanglement is the name for a condition of a multi-party quantum state, a condition that exists jointly among the parties. Such a state allows the participating parties to exhibit properties that are often labeled counter-intuitive, and some of these were brought to wide attention by the publication of an article by Albert Einstein, Boris Podolsky and Nathan Rosen in 1935 \cite{EPR}. Their article became the center of a sufficiently controversial discussion that it is now widely known by the initials of the three authors - EPR. 

Erwin Schr\"odinger \cite{schr1,schr2} was the physicist who introduced the term ``entanglement" for quantum states while responding to EPR. He did this in two articles in 1935, in one of which he introduced his famously entangled Cat, while illustrating entanglement's role in the organization of quantum states belonging jointly to two quantum systems. 

As Schr\"odinger noted, entanglement is an interesting concept mathematically, {\it e.g.}, in connection with the Schmidt theorem of analytic function theory \cite{schmidt1907,fedorov2014}, and in this way has a side-door connection with the singular value decomposition theorem for matrices. Entanglement attracts seriously focused attention in physics because it is a key resource in quantum information processing, and is needed for exotic quantum processes such as teleportation \cite{bennett1999}. 

It is surprising now that, more than 80 years later, a challenge still exists to describe how to take a satisfactory measure of entanglement. That specific challenge is addressed here, showing how to go beyond known measures for states of two entangled parties such as Bell states, and tell how much entanglement resource is present in a given multi-party joint state. Progress in this direction was slow for decades, at the same time that physical devices making use of multi-party quantum entanglements were being regularly operated in labs around the world. That specific challenge has now been overcome \cite{xie2021}, as we will explain, but only for three-party  entanglement. However, that recent advance suggests a route for still further progress. 

There are essential points to be remembered. The state of a quantum system exists only as a vector in an abstract mathematical vector space, a space assigned to a single degree of freedom (DOF) of the quantum system. Moreover, a quantum state doesn't contain information about its system in the definite way familiar to most physical scientists. That is, quantum states yield knowledge about physical systems not in terms of facts, but in terms of probabilities, even, one could say, in the form of square roots of probabilities, since Born's foundational interpretation of quantum theory is a rule that brings probability into view only via the absolute square of the state itself. 

We will first review briefly how information in the form of probability is successfully captured by entanglement by giving examples using quantum systems called qubits (pronounced ``kew-bits" in English). 

It is helpful to keep in mind that qubit is a well-defined term for a real physical system. More precisely it's the name for just a single degree of freedom of a real physical system. There is a further restriction on the name: the term qubit is applicable not only to one of a system's degrees of freedom, but even then only if the vector space of that degree of freedom has just two dimensions. For example, a qubit cannot represent an electron. An electron has too many degrees of freedom. 

But a qubit can represent one of an electron's degrees of freedom, such as its two-valued spin-$\frac{1}{2}$ character. We can say the states of electron spin are the two states of a qubit and write those states $|0\ra$ and $|1\ra$, or as $|0\ra_A$ and $|1\ra_A$ when we want to show the subscript $A$ as the label for the specific degree of freedom as, e.g., the spin of a specified electron. The qubit's two states $|0\ra$ and $|1\ra$ then contain all the information available about that spin degree of freedom. The 0 and 1 appearing in the  (Dirac notation) brackets are deliberately used so as to indicate that $|0\ra$ and  $|1\ra$ are quantum analogs of the distinct 0 and 1 states that are possible for a classical ``bit" of information.

The term distinct means here that $|0\ra$ and  $|1\ra$ are orthogonal vectors, i.e., that the scalar product (inner product) of the two vectors vanishes: $\la 0|1\ra = \la 1|0\ra = 0$. The two states are also normalized: $\la 0|0\ra = \la 1|1\ra = 1$. This allows them to be used as basis states in their space. In this way an abstract vector space can be used to represent mathematically any value of two opposite states of a physical degree of freedom (again, e.g., an electron's spin-half or, as another frequent example, the  possible values of the two opposing transverse polarizations of a photon). 

The two states of a qubit can also represent intermediate possibilities between the 0 and 1 states by using vector superposition. A key point to repeat is that independent states of a two-valued degree of freedom cannot exist simultaneously in the classical world of a single system (say the head and tail of a coin being flipped), whereas their quantum analogs are actually present together, at once, as probabilistic possibilities, in every non-zero two-state superposition such as this for the ``opposite" (orthogonal) quantum basis states $|0\ra$ and  $|1\ra$: 
\begin{equation} \label{qubit}
|\psi\rangle = a|0\rangle + b |1\rangle,
\end{equation}

We emphasize that, between the numbers 0 and 1 and the pair of brackets $|0\rangle$ and $|1\rangle$ assigned to a quantum state, the key difference first appears when a qubit is in a state of vector superposition of basis states $|0\ra$ and  $|1\ra$, as in (\ref{qubit}) above. This is a familiar property of waves of all kinds, and in this way qubit states are given a wave-like capability. This was sensed by Einstein already in 1923 to be the essential point of Louis de Broglie's revolutionary ``duality" concept. Einstein's notes to Born and to Langevin about de Broglie's proposal have been quoted as saying ``Read it. Even though it might look crazy, it is absolutely solid." and ``He has lifted a corner of the great veil." Recall that 1923 was several years before Schr\"odinger later discovered the wave equation that is obeyed by de Broglie's wave concept. To repeat, the wave-like superposition of independent and opposite states of a degree of freedom is available to qubits.

The freedom of the values to be assigned to the complex numbers $a$ and $b$ in (\ref{qubit}) allow $|\psi\ra$ to be any state from $|0\ra$ to its orthogonal partner $|1\ra$. We are here accepting that all quantum states, including superpositions such as $|\psi\ra$, are under the restriction of the Born probability rule and must maintain the state's scalar product (unit norm) condition $$\la\psi|\psi\ra = |a|^2 + |b|^2 = 1.$$ 

\section{Entanglement and Concurrence Appear}
Going further, there is no bar to considering two-party tensor products of the states for party A and party B in two-qubit superposition states such as this:
\begin{equation}\label{super1}
|\phi_1\ra_{AB} = \Big(a|0\ra_A + b |1\ra_A\Big) \otimes \Big(a'|0\ra_B + b'|1\ra_B\Big).
\end{equation}
The vector product symbol $\otimes$ between $A$'s state and $B$'s state, points to the tensor nature of the difference between the $A$ state space and the $B$ state space. The labeling of states by names $A$ and $B$ is enough to assign correct vector-space memberships to the different members of vector products, so the symbol $\otimes$ will frequently be omitted.

Note also that the states of knowledge expressed by $|1\rangle_A$ and $|0\rangle_B$ are two independent elements of knowledge, something that we should keep in mind. The novelty of their independence increases dramatically in products of superpositions of two-party states such as in (\ref{super1}), {\bf because in (\ref{super1}) we first encounter entanglement} but in the sense that it is absent. That is, we declare: 
\begin{center}
{\em A two-party $AB$ qubit product state, such as (\ref{super1}), is not entangled}. 
\end{center}
Our first expression of the definition of entanglement is thus in reverse, by defining that non-entanglement refers to simple tensor-multiplication to create a single joint-state $AB$ that is a product. Different descriptive terms in addition to ``two-party product state", such as ``separable state" or ``factorizable state" are also used for this form.

Today, more than eight decades after Schr\"odinger named it, and despite that elementary mathematical foundation for its definition, entanglement prompts intense discussions among physicists and natural philosophers concerning connections between entanglement and the facts of multi-particle reality. In spite of serious ongoing discussions of philosophical points, there are no doubts about entanglement itself, in the sense that there is complete acceptance that extremely strange processes exist that depend on it, processes ultimately compatible with de Broglie's duality and Born's probability postulate. A good example is teleportation, described by Bennett {\it et al.} in 1993 \cite{bennett1993} and experimentally recorded \cite{bouwmeester1997} and brought under laboratory control since then. Such processes are impossible without the inclusion of entanglement. They have made entanglement a centerpiece in the current worldwide race toward both scientific and commercial exploitation of quantum processes, foretelling significant improvements in speed and security of communication and computing \cite{gottesman1999,arute2019,ekert1991}.

Specifically, in the teleportation process, in order for Alice to send one qubit secretly to a distant receiver Bob, they have to share one unit of entanglement resource (sometimes called one ``ebit"), which is contained in any one of the four following joint states:
\begin{equation}\label{4Bell}
\begin{split}
&|\Phi^{\pm}\rangle_{AB} =\frac{1}{\sqrt{2}}\Big(|0\rangle_A\otimes|0\rangle_B \pm|1\rangle_A\otimes|1\rangle_B\Big),\\
&|\Psi^{\pm}\rangle_{AB} =\frac{1}{\sqrt{2}}\Big(|0\rangle_A\otimes|1\rangle_B \pm|1\rangle_A\otimes|0\rangle_B\Big),
\end{split}
\end{equation}
commonly called Bell states. These four states were found to contain the largest amount of entanglement resource. If a different entangled state were chosen instead, the performance of teleportation would be reduced, signaling a smaller amount of entanglement resource contained in that state.

The fact is, considered as a resource, entanglement not only answers a yes or no question about a state (entangled or not), but also  assigns a quantitative value, a degree of entanglement. The existence of such an {\it entanglement measure}, elevates entanglement from an interesting fundamental concept to a useful component in performing practical tasks \cite{chitambar2019}. Specific open questions are then obvious: if an entangled state is on hand, how much entanglement resource is present? Or first, how does one find a measure to quantify entanglement? 

One can learn quickly that this might not be a simple matter. For example, what if a two-party state includes more than a single product, as below in (\ref{pqrs})? Here there is a sum of four distinct products for parties $A$ and $B$:
\begin{equation}\label{pqrs}
|\psi_{pqrs}\ra_{AB} = p|0\ra_A |0\ra_B + q|0\ra_A |1\ra_B + r|1\ra_A|0\ra_B + s|1\ra_A|1\ra_B,
\end{equation}
where $p$, $q$, $r$, $s$ are just complex numerical coefficients. The state $\eqref{pqrs}$ is written as a superposition of the basis vectors of the joint $AB$ space. We must properly normalize it in conformity to the Born rule, requiring $\la \psi_{pqrs}|\psi_{pqrs}\ra = 1$. The orthonormality of the one-party basis states $|0\ra$ and $|1\ra$ quickly shows that $|p|^2 + |q|^2 + |r|^2 + |s|^2 = 1$ is Born's requirement. 

We can ask: could this sum of four terms in (\ref{pqrs}), despite its appearance, nevertheless just be a product state? A general two-qubit product state can always be written in the form \eqref{super1}, so its products can simply be multiplied out, which gives another equivalent form of $|\phi_1\rangle_{AB}$:
\begin{equation}\label{super1mult}
|\phi_1\rangle_{AB} \to |\phi_{1,mult}\ra_{AB} = aa'|0\ra_A|0\ra_B + ab'|0\ra_A|1\ra_B + ba' |1\ra_A|0\ra_B + bb'|1\ra_A|1\ra_B.
\end{equation}

A direct observation is this: If the general two-qubit $pqrs$ state \eqref{pqrs} is a product state, there must be some specific values of $a,a',b,b'$ such that $p=aa',\ q=ab',\ r=a'b,$\  and\ $s=bb'$ hold. Thus, $ps=qr$ must be true, since they both equal $aa'bb'$. We next show that the reverse statement is still true, that is: If a general two-qubit $pqrs$ state \eqref{pqrs} has the property that $ps=qr$, we immediately know that it is a product state.

To see this, one groups the four terms in \eqref{pqrs} and factors out both $|0\rangle_B$ and $|1\rangle_B$, which leads to
\begin{equation}\label{pqrs2}
  |\psi_{pqrs}\rangle_{AB}=\Big(p|0\rangle_A+r|1\rangle_A\Big)|0\rangle_B+\Big(q|0\rangle_A+s|1\rangle_A\Big)|1\rangle_B.
\end{equation}
It is obvious that if $ps=qr$, or equivalently $p/r=q/s$, the $A$ party states in the two parentheses are the same, i.e., differing only by an irrelevant factor. One can then immediately conclude that \eqref{pqrs2} is a product state.

What do we learn from this lesson? It's clear that a {\em difference} between $ps$ and $qr$ serves as a measure of the {\it inability} to express the state \eqref{pqrs} as a product state. It's a measure of the distance between that general state and the condition of product states. 

This is what we want because our entanglement definition states that it is departure of a given state from the product form that creates entanglement! This observation was grasped by William Wootters, who created the qubit entanglement measure $C$ named {\em Concurrence}. It followed his examination of the requirements for what is called entanglement of formation, with his colleagues two decades ago \cite{hill1997}. For the pure two-qubit state $|\psi_{pqrs}\ra$ one has this simple formula for Wootters' qubit entanglement measure called Concurrence:
\begin{equation} \label{C}
    C(|\psi_{pqrs}\rangle) = 2|ps - qr|, \quad {\rm where}\quad 1 \ge C \ge 0.
\end{equation} 
The bounds between $1$\ and $0$\ follow from Born's condition $|p|^2 + |q|^2 + |r|^2 + |s|^2 = 1$. 

For our general two-party qubit state $|\psi_{pqrs}\rangle$ a numerical result for $C$ is quickly obtained by returning to the specific choice made in our stated example: $p = aa', q = ab', r = a'b,$ and $s = bb'$. These lead to $ps = aa'bb' = qr$, and so $C(pqrs) = 0$, thus conforming to what we already knew about the coefficient relations, that $|\psi_{pqrs}\ra$ originated in product state $|\phi_1\ra_{AB}$, and so it could not be an entangled state, making $C = 0$ inevitable.

Another application of the Concurrence formula is worth mentioning because the well-known Bell states \eqref{4Bell} used for teleportation are examples of the ``$pqrs$" type. 
The Bell states can serve as new two-party $AB$ basis states. The reader can easily check that our $pqrs$ state \eqref{pqrs} can be expressed by the new basis as
\begin{equation}
|\psi_{pqrs}\rangle =  \Big(\dfrac{p+s}{\sqrt{2}}\Big)|\Phi^+\rangle_{AB}+ \Big(\dfrac{q+r}{\sqrt{2}}\Big)|\Psi^+\rangle_{AB}+ \Big(\dfrac{q-r}{\sqrt{2}}\Big)|\Psi^-\rangle_{AB}+ \Big(\dfrac{p-s}{\sqrt{2}}\Big)|\Phi^-\rangle_{AB}
\end{equation}
It is another recommended simple exercise to show that Concurrence is maximal ($C = 1$) for each Bell state.  Because Bell states are not product states, one cannot separate their $A$ qubit states from their joint $B$ qubit states and speak of an $A$ character alone. What's more, Bell states are maximally entangled ($C = 1$), which conforms our expectation when they are applied to the teleportation process.

To be more specific, when the system $AB$ is in Bell joint state $|\Phi^+\ra_{AB}$, as given in (\ref{4Bell}) above, the qubit $A$ is NOT just in the $A$ state $|0\rangle_A$, and NOT just in the $A$ state $|1\rangle_A$, even though both $|0\rangle_A$ and $|1\rangle_A$ are present, and NOT even in any possible single-qubit-superposed state for $A$ such as $a|0\rangle_A + b|1\rangle_A$. 

Instead, one can directly ask what information does $|\Phi^+\ra_{AB}$ contain about the state of $A$ by evaluating the inner product of $|\Phi^{+}\ra_{AB}$ with the two $A$ states: $_A\la 0|$ and $_A\la 1|$, which yield the results
\begin{equation}
    \begin{split}
        _{A}\la 0|\Phi^+\ra_{AB} = \dfrac{1}{\sqrt{2}}|0\ra_B,\quad {\rm and}\quad 
       _{A}\la 1|\Phi^+\ra_{AB} = \dfrac{1}{\sqrt{2}}|1\ra_B.
    \end{split}
\end{equation}

One immediately sees two features of significance. First, the two possible outcome states are not normalized, and second, in both cases the result for $A$ retains dependence on $B$. In fact, according to Born's rule of normalization, the norms of the outcome states provide residual probabilities, and in this case, the norms of the two inner products share probabilities equalling one-half. Additionally, the dependences on $B$ that remain are correctly reminding us of a property of the original state $|\Phi^+\ra_{AB}$: one obtains  zero probability immediately if one asks for information in $|\Phi^+\ra_{AB}$ about $|0\ra_A$ and $|1\rangle_B$ simultaneously or about simultaneous $|1\ra_A$ and $|0\ra_B$. 

One should note a consistency. The probabilistic features of the measurement do not rely on which orthogonal vectors one chooses to represent party $A$ at the beginning. Suppose that the two superposed states $|0'\rangle_A\equiv\Big(|0\rangle_A+|1\rangle_A\Big)/\sqrt{2}$ and $|1'\rangle_A\equiv\Big(|0\rangle_A-|1\rangle_A\Big)/\sqrt{2}$ are chosen instead of $|0\ra_A$ and $|1\ra_A$. The result then becomes
\begin{equation}
    \begin{split}
        \la 0'|\Phi^+\ra_{AB} = \dfrac{1}{\sqrt{2}}|0'\ra_B,\quad {\rm and}\quad 
       _{A}\la 1'|\Phi^+\ra_{AB} = \dfrac{1}{\sqrt{2}}|1'\ra_B.
    \end{split}
\end{equation}
and the same probabilistic features remain.

Naturally we also want to study how interactions (usually introducing decoherence losses, for instance) can affect entanglement, specifically to study entanglement in dynamical situations. This kind of interaction studies can lead to unexpected results, and examples are available. For example, Yu and Eberly noticed \cite{yu2004} that when subject to interaction with the vacuum, two-qubit entanglement (in their case a pair of two-level atoms in two distinct cavities with atom states entangled), can vanish abruptly, after only a finite time following initiation, instead of decaying forever smoothly in the expected exponential way. The abruptness is known as ``entanglement sudden death'' \cite{yu2009}. In addition, one of our more recent results \cite{ding2021} showed that entanglement among a three-qubit system sharing a single excitation can remain at a constant value for a non-zero finite time, which can be termed ``entanglement sudden freezing''.

Last, we note that although Concurrence provides an entanglement measure to quantify the degree of two-qubit entanglement in quantum states, it is not the only such measure. Other examples include  entanglement of formation \cite{bennett1996}, Negativity \cite{vidal2002} and normalized Schmidt weight \cite{grobe1994}. A fundamental result is that all such two-qubit entanglement measures are equivalent in the sense that they give the same answer to the question whether one specified state is more entangled than another \cite{singh2020}. In this sense, it is sufficient to consider only one entanglement measure in most two-qubit scenarios.

\section{Counter-intuitive or Not?}
The often-remarked ``counter-intuitive" nature of entanglement commonly arises because an action on party $A$ of an $AB$ entangled state is sometimes said to lead to an instantaneous ``reaction'' by a distant $B$ (e.g., an immediate change of state by party $B$). 

As a specific example we can consider the two-photon pair generated by the spontaneous parametric down conversion process \cite{huang1993}. The polarization degrees of freedom for the $A$ and $B$ photon pair, are entangled in the form of $|\Phi^+\rangle_{AB}$ as in \eqref{4Bell}. We then send photon $A$ to Alice, and photon $B$ to a distant receiver Bob. By saying ``send'', we mean that we continuously change the two photons' physical locations. The locations are obviously two independent degrees of freedom different from the polarizations belonging to $A$ and $B$. Therefore, if the experimental setup is ideally prepared, one can assume that the polarization states are preserved throughout the sending process, obtained as  $|\Phi^+\rangle_{AB}$ via the down conversion process, even though the two photons are sent to remote locations.

Meanwhile, the state of either $A$ alone or of $B$ alone is a mystery even if the joint-state of the $AB$ system is well known to be a Bell state. We suppose that the $A$ observer, Alice, makes a measurement to query the polarization of the $A$ photon. As was explained above, Alice's measurement process allows two distinct operations and each corresponds to a distinct outcome. Suppose the outcome turns out that the projection $_{A}\langle0|$ is applied to the original state $|\Phi^+\rangle_{AB}$. Then immediately Alice knows the joint state becomes $|0\rangle_A|0\rangle_B$. In this way, and counter-intuitively, the state of a (possibly remote) degree of freedom (the polarization of $B$) undergoes an immediate change due to Alice's instant knowledge, coming from the joint state's coupling with the projection action on $A$. This was termed ``spooky action at a distance'' by Einstein.

However, we recall that a quantum state is intended to convey the amount of information available about a degree of freedom. It is obvious that the information about $B$'s polarization after the measurement is only possessed by Alice. If Alice is considering the consequences of the measurement, in order to bring its consequence to Bob's attention, then the distance between their supposed physical locations prevents her from doing so instantly. In fact, Alice has to apply ordinary transmission channels ({\it e.g.}, imagine a laser beam signal), and the speed of the channel cannot be faster than light speed.\\

\section{Qubit Power for More than Two Qubits}

We have mentioned that studies of multi-party entanglement beyond Bell states have been incomplete and thus frustrated for many decades. However, it has recently become clear that Concurrence can continue to be useful in a much larger domain when it leads to exploitation (see below) of what can be called the ``Qubit Power" of the Schmidt theorem \cite{schmidt1907,fedorov2014}. One form of the Schmidt theorem deals with states of an arbitrarily large number of qubits, which we can write as
\begin{equation}\label{n-qubit pure state}
|\Psi _{1,2,\cdots,n}\rangle =\sum_{s_{1},s_2,\cdots,s_{n} = \{0,1\}} c_{s_1,s_2,\cdots,s_n}|s_{1}\ra|s_{2}\ra\cdots|s_{n}\rangle,  
\end{equation}%
where $c_{s_{1},s_2,\cdots,s_{n}}$ are normalized coefficients and $s_{j}$
takes values 0 or 1 corresponding to the available states $|0\rangle $,
$|1\rangle $ of the $j$-th qubit, with $j=1,2,\cdots,n$. 

The Schmidt theorem proves the following remarkable fact: when a single qubit $A$ is in a joint state with other qubits $\{B,C,D,\cdots\}$ as in Eq. \eqref{n-qubit pure state}, the multi-qubit combination $\{B,C,D,\cdots\}$, even when representing a much larger DOF beyond $A$, behaves just like one two-state qubit. More specifically, despite the large dimension of the combined DOFs in $\{B,C,D,\cdots\}$, only two ``distinct'' states are ``active'' with $A$. All the others are suppressed by the presence of qubit $A$.

As a simple example, we consider the following tripartite state called the $W$ state, conventionally written as:
\begin{equation}\label{wstate}
    |W\rangle_{ABC}=\dfrac{1}{\sqrt{3}}\Big(|100\rangle_{ABC}+|010\rangle_{ABC}+|001\rangle_{ABC}\Big).
\end{equation}
The $\{BC\}$ sub-combination can be considered as a larger single DOF with 4 dimensions, and its simplest choice of basis vectors is $\{|00\rangle_{BC},\ |01\rangle_{BC},\ |10\rangle_{BC},\ |11\rangle_{BC}\}$. Another way to choose the basis vectors, more convenient to analyze the $W$ state, is given by 
\begin{equation}
\begin{split}
    &\{|00\rangle_{BC},\quad \frac{1}{\sqrt{2}}\big(|01\rangle_{BC}+|10\rangle_{BC}\big),\quad \frac{1}{\sqrt{2}}\big(|01\rangle_{BC}-|10\rangle_{BC}\big),\quad |11\rangle_{BC}\}\\
    &\equiv\{ |0'\rangle_{BC},\ \ |1'\rangle_{BC},\ \ |2'\rangle_{BC},\ |3'\rangle_{BC}\}.
\end{split}
\end{equation}

The new basis vectors can easily be checked to be orthonormal. When expressed in these basis vectors, the $W$ state becomes
\begin{equation}\label{wstate2}\begin{split}
    |W\rangle_{ABC}\ =\ &\dfrac{\sqrt{2}}{\sqrt{3}}\ |0\rangle_{A}\left[\dfrac{1}{\sqrt{2}}\Big(|01\rangle_{BC}+|10\rangle_{BC}\Big)\right]+\dfrac{1}{\sqrt{3}}\ |1\rangle_{A}|00\rangle_{BC}\\
    =\ &\dfrac{\sqrt{2}}{\sqrt{3}}\ |0\rangle_{A}|1'\rangle_{BC}+\dfrac{1}{\sqrt{3}}\ |1\rangle_A|0'\rangle_{BC}.
\end{split}
\end{equation}
In \eqref{wstate} and \eqref{wstate2}, we quickly recognize what we are calling the ``Qubit Power'' of the Schmidt theorem. We see that $|0\rangle_A$ and $|1\rangle_A$, the two states of qubit $A$, can combine with only $|0'\rangle$ and $|1'\rangle$ of the $BC$ pair. That is, by coupling it to qubit $A$, the four-dimensional $BC$ joint state is reduced to act as another two-dimensional qubit. The $W$ state then behaves as a two-qubit state and the Concurrence expression \eqref{C} can now be easily applied and gives $C = 2\sqrt{2}/3$. Naturally, it must be understood as $C_{A(BC)}$, the Concurrence between $A$ and the pair $(BC)$. The same argument can also be applied when we separate qubit $B$, or qubit $C$, and for $W$ we achieve the same result: $C_{B(CA)}=C_{C(AB)}=2\sqrt{2}/3$. For a generic three-qubit system, when considering the entanglement between one qubit and the remaining two taken together as an “other” single party, the three are naturally called ``one-to-other'' bipartite entanglements.

\section{Genuine Entanglement for Three Qubits}
The ``Qubit Power'' resulting from the Schmidt decomposition extends two-party entanglement to multi-party systems. We can continue this exploration and demonstrate a way to quantify a new type of three-party entanglement that doesn't exist for two-party systems. Drawing on our experience with two parties in Sec. 2, we begin with an easily expected parallel statement for three parties: 

\begin{center}
{\it Three-party product states have zero tripartite entanglement.}
\end{center}
This clearly leaves open the obvious ``how much'' question for any three-party state that is not a product state. That is, how are we to quantify the degree of more general tripartite entanglements with a specific measure?

Unfortunately, none of the one-to-other concurrences following directly from Qubit Power are able to exhibit needed three-party features. Proposals have been made but satisfactory progress began in various stages only after attention was focused on the ground-breaking analysis of D\"ur, Vidal and Cirac \cite{dur2000} who identified the classes of ``genuinely'' entangled states in three-qubit systems.  This prompted a focus on just three-party entanglement as an example of what is now called ``genuine multipartite entanglement'' and labeled GME. 

Specifically, in D\"ur, Vidal, and Cirac \cite{dur2000}, all three-qubit states were able to be separated into four distinct classes:\\ 

\noindent (i) The product states of the form $|\psi\rangle_{ABC}=|\alpha\rangle_A|\beta\rangle_B|\gamma\rangle_C$, which is certainly the simplest class since its members can have no entanglement.\\

\noindent (ii) The biseparable states (also called ``one-to-other" separable states) are those of the form $|\psi\rangle_{ABC}=|\alpha\rangle_A|\phi\rangle_{BC}$, where one of the qubits (here qubit $A$) is separated out by tensor factoring. Entanglement exists between qubits $B$ and $C$, and therefore, entanglement in such tripartite states is not GME. \\

\noindent (iii and iv) Two distinct classes are genuinely entangled, the Greenberger-Horne-Zeilinger (GHZ) class and the $W$ class \eqref{wstate}. Those two individual states are themselves the most entangled states in each of those classes respectively. The $W$ state is given in \eqref{wstate} and the GHZ state is
\begin{equation}\label{ghzstate}
    |\text{GHZ}\rangle_{ABC}=\dfrac{1}{\sqrt{2}}\Big(|0\rangle_A|0\rangle_B|0\rangle_C+|1\rangle_A|1\rangle_B|1\rangle_C\Big).
\end{equation}
D\"ur, Vidal, and Cirac identified features of those key state classes, such as follows: The GHZ state is the maximally entangled state of three qubits, but if one of its three qubits is traced out, the remaining state is completely disentangled. Thus, the entanglement properties of states in the GHZ class are very fragile under particle loss. Oppositely, the entanglement of $W$ is still present under disposal of any one of its three qubits.

The definition of tripartite entanglement must be modified to include the ``genuine'' property. This means accepting two specific conditions, (a) and (b), identified by Ma {\it{et al.}} \cite{ma2011}, and a third condition (c) appearing later \cite{xie2021}:\\[0.7em]
\noindent (a) {\it Genuine tripartite entanglement is absent for three-qubit product states and biseparable states.}\\

\noindent (b) {\it Genuine tripartite entanglement is present for three-qubit non-biseparable states (all those of GHZ class and $W$ class).}\\

\noindent (c) {\it ) Genuine tripartite entanglement measure ranks the GHZ state
as more entangled than the W state.}\\ [0.7em]
The introduction of the ``genuine'' concept, which labels biseparable states as not genuinely entangled, has practical justification. An important background fact is that there are quantum tasks for more than two parties that need genuine entanglement to succeed. As an example, three-party-assisted teleportation can be expected to work reliably if and only if the entangled state engaged and shared by three parties (Alice, Bob, and Charlie), is genuinely entangled \cite{karlsson1998}. Thus, genuine tripartite entanglement can be considered as a three-party common resource in practical quantum tasks.

The three new conditions (a) and (b) and (c) for genuine entanglement make the measurement of multipartite entanglement complicated but interesting. A series of multipartite measures were previously invented in the two recent decades and developed, but most of them are not genuine. On the one hand, examples are multipartite monotones by Barnum and Linden \cite{barnum2001}, a Schmidt measure $P$ by Eisert and Briegel \cite{eisert2001, hein2004}, and global entanglement $Q$ by Meyer {\it et al.} \cite{meyer2002,brennen2003}. All of these, as well as generalized multipartite concurrence $C_N$ by Carvalho {\it et al.} \cite{carvalho2004}, fail to satisfy condition (a). On the other hand, the famous 3-tangle by Coffman {\it et al.} \cite{coffman2000,miyake2003}, as well as entanglement based on ``filters'' by Osterloh and Siewert \cite{osterloh2005}, a GME based on a positive-partial-transpose (PPT) mixture by Jungnitsch {\it et al.} \cite{jungnitsch2011}, and the multi-party coherence advanced by Qian {\it et al.} \cite{qian2020} violate condition (b). There are also several measures based on identifying the distance between a given state and its closest product state (see examples in \cite{shimony1995, plenio2001, wei2003}). From their definitions, they violate condition (a). 

Our recent work \cite{xie2021} has been successful in meeting the challenging GME problem for three-qubit systems by extending the use of Concurrence in a novel manner that does satisfy all requirements (a) and (b) and (c). The measure uses a surprising geometric approach, as we explain next.\\

\section{Triangles and GME Measures}
To construct a GME measure correctly for a three-qubit system, a genuinely entangled three-qubit state requires all three one-to-other bipartite concurrences to be positive: $C^2_{A(BC)}>0,\ C^2_{B(CA)}>0,$ and $C^2_{C(AB)}>0$. The reverse is also true. Thus, it is attractive to relate a genuine tripartite entanglement measure to these three quantities in a new way, which was found to be possible. 

\begin{figure}[t]
    \centering
    \includegraphics[width=0.3\textwidth]{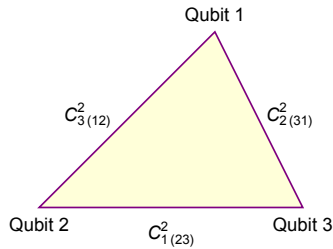}
    \caption{The concurrence triangle for a three-qubit system. The square of the three one-to-other bipartite concurrences are equal to the lengths of the three edges.}
    \label{fig:triangle}
\end{figure}

Importantly, it had been proved earlier by Qian, Alonso, and Eberly that the three one-to-other entanglements are not completely independent \cite{qian2018}. In their work using the concurrence measure, their \textit{entanglement polygon inequality} states that one of the three one-to-other entanglements cannot exceed the sum of the other two, as follows:
\begin{equation}
    C_{A(BC)}\leq C_{B(CA)}+C_{C(AB)}.
\end{equation}
The same relation was also proved \cite{qian2018} to be applicable beyond Concurrence to negativity, von Neumann entropy, and normalized Schmidt weight. A stronger version was proposed by Zhu and Fei in \cite{zhu2015}, where all three concurrences are replaced by their squared forms,
\begin{equation}\label{trianglerelation}
    C_{A(BC)}^2\leq C_{B(CA)}^2+C_{C(AB)}^2.
\end{equation}
An obvious polygon interpretation \cite{qian2018} for both of these inequalities is that the three squared (or not) one-to-other concurrences can represent the lengths of the three sides of a triangle. When referred to the squared formula \eqref{trianglerelation}, we call it the {\it concurrence triangle}. Any given three-qubit pure state has a unique concurrence triangle, illustrated in Fig. \ref{fig:triangle}.

The concurrence triangle can have various shapes for different states. For the simplest three-qubit product states, all concurrences vanish: the concurrence triangle has all its three edges as zero, and is then a single dot, and there is no entanglement.  For a biseparable state in a three-qubit system, suppose the qubit $A$ is singled out. Then we have $C^2_{A(BC)}=0$, but $C^2_{B(CA)}>0$ and $C^2_{C(AB)}>0$. In this way, one of the edges is zero. Two vertices of the concurrence triangle coincide, so the triangle is a line. The most complicated situation is when the state is genuinely entangled. All the three edges are then positive. But the concurrence triangle can still have two different shapes: the three vertices are collinear or they span a plane.

Remarkably, the results of some earlier attempts to find 3-party entanglement measures can also be described in terms of a triangle. One example is the Global Entanglement labeled $Q$ by Meyer {\it et al.} \cite{meyer2002,brennen2003}. Its value is a numerical multiple of the perimeter of the concurrence triangle. By a close check, one sees that Global Entanglement assigns correct values to product states and genuinely entangled states. However, there is no genuine entanglement for biseparable states, but the concurrence triangles for these states have positive perimeters. This indicates that Global Entanglement violates condition (a) by assigning wrong values to the biseparable class, and thus it cannot measure GME. 

Additionally, Ma {\it et al.} proposed another measure called {\it Genuine Multipartite Concurrence} (labelled GMC) as $\text{min}\Big\{C^2_{1(23)},C^2_{2(31)},C^2_{3(12)}\Big\}$ \cite{ma2011}. That measure was later extended by Hashemi-Rafsanjani  \cite{rafsanjani2012} who used $X$-form mixed states \cite{yu2007} to find an explicit expression for it. The GMC measure has its own triangle-geometric meaning: the length of the {\em shortest edge} of the concurrence triangle. With a quick check, one finds that GMC assigns correct values to all classes of states. GMC then satisfies both condition (a) and (b) of Ma, et al, and is a tripartite GME measure. Along a similar line, in our recent work \cite{xie2021}, we proved that the {\em area} of the concurrence triangle is also a GME measure. It is zero for both product and biseparate states, and thus satisfies condition (a) for GME. Figure 2 shows a table of relations between concurrence triangles and the required values, zero or non-zero, for triangle features such as area and side lengths.

\begin{figure*}[t]
\centering
    \includegraphics[width=0.9\textwidth]{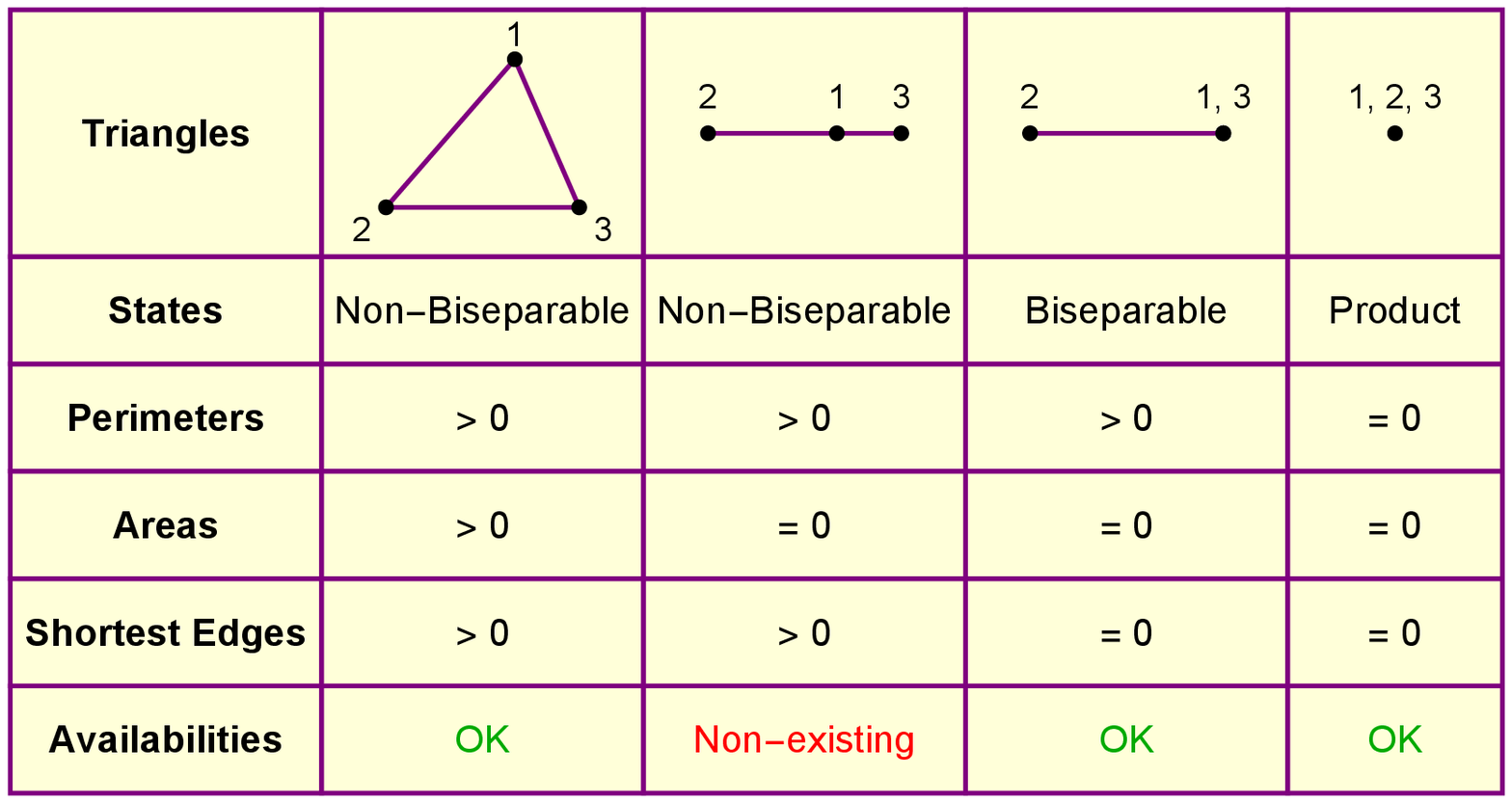}
    \caption{The table of possible states and their corresponding concurrence triangles. The values of three multipartite entanglement measures (perimeters for Global Entanglement, areas for Concurrence Fill and shortest edges for GMC) are compared with zero. One class of the non-biseparable triangles is proved to be non-existing.}
    \label{fig:table}
\end{figure*}

Concern may arise from the fact that genuinely entangled states have two different forms of concurrence triangle, one of which has zero area (when the three vertices are collinear). In this way, this measure would assign 0 value to these genuinely entangled states, and so violates condition (b). But this is inapplicable because of a surprising fact we discovered: all genuinely entangled states have positive areas for their concurrence triangles. Generically, a triangle has zero area when its three vertices are collinear. Our discovery excludes the possibility that the three vertices are collinear while no two vertices coincide, which is a condition corresponding to the nonbiseparable states. This discovery guarantees that the {\em area} of a concurrence triangle also satisfies condition (b), and we gave it a name: {\it Concurrence Fill}. Heron's classic formula for triangle area in terms of side lengths leads to the following explicit expression for our triangle measure denoted $F_{123}$:
\begin{eqnarray}\label{heron}
        &F_{123}\equiv\left[\dfrac{16}{3}Q\left(Q-C^2_{1(23)}\right)\left(Q-C^2_{2(13)}\right)\left(Q-C^2_{3(12)}\right)\right]^{1/4},\nonumber\\
        &\text{where}\quad Q=\dfrac{1}{2}\left(C^2_{1(23)}+C^2_{2(13)}+C^2_{3(12)}\right).
\end{eqnarray}
$Q$ is the half-perimeter (and thus equivalent to Global Entanglement), while the prefactor $16/3$ is for normalization.

\section{Discussion, Extensions, and Summary}
In summary, we introduced the background history and the challenging nature of quantum entanglement. We recalled that the state of a quantum physical system exists only as a vector in an abstract mathematical vector space. We introduced qubit as a term to identify the two-dimensional vector assigned to a two-valued degree of freedom of a real physical system, such as the spin-half values available to an electron but not the entire electron. The qubit's state vector conveys all available information about the degree of freedom. We have been using the symbols $|0\ra$ and $|1\ra$ as orthonormal basis vectors in the space assigned to the degree of freedom of interest, and have discussed combinations such as superpositions and also products of superpositions of those basis vectors when two physical systems (degrees of freedom) labeled $A$ and $B$ are available:
\begin{equation}\label{super1'}
|\phi_1\ra_{AB} = \Big(a|0\ra_A + b |1\ra_A\Big) \otimes \Big(a'|0\ra_B + b'|1\ra_B\Big).
\end{equation}

We identified a state's entanglement with the failure of the state to take product form, and showed that this provides a route to understanding the quantity called Concurrence as a reliable measure of two-qubit entanglement. Then we introduced the features that have made it challenging for several decades to extend that measure beyond 2-qubit entanglement, and finished with a current understanding that solves the 3-qubit entanglement challenge. 

One of the most significant difficulties is the extremely high dimensionality of the Hilbert space for multi-qubit systems, which grants the possibility of having inequivalent multipartite entanglement measures (see Vidal [2]). This complication implies the need to divert attention from entanglement to the study of multi-qubit Hilbert space itself. Specifically, we recalled a fundamental division for 3-qubit state space by D\"ur {\it et al.} \cite{dur2000}, where two completely distinct genuinely entangled classes were identified as the GHZ class and the $W$ class. This identification helps to constitute three conditions for a three-party genuine entanglement measure. 

A geometric interpretation of symmetric combinations of multi-party one-to-other Concurrences was found by Qian {\it et al.} \cite{qian2018} in terms of polygons. Surprisingly, we were able to exploit this by first defining a specific concurrence triangle for three-qubit systems, as illustrated in Fig. \ref{fig:triangle}, and then showed that the concurrence triangle's area fulfills all of the necessary conditions to become a fully satisfactory GME (genuine multipartite entanglement) measure for three qubits. Detailed information of concurrence triangle and tripartite measures was summarized in the table in Fig. \ref{fig:table}. 

\begin{figure}
    \centering
    \includegraphics[width=0.45\textwidth]{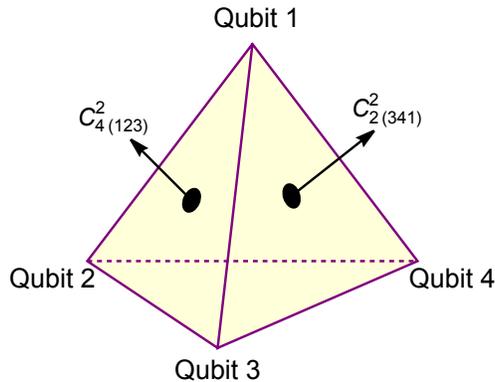}
    \caption{The Concurrence Tetrahedron, a possisible candidate of the geometric solution for 4-qubit GME measure. The square of the four one-to-other bipartite concurrences are equal to the areas of the four surfaces.}
    \label{fig:tetrahedron}
\end{figure}

Now the nature of open questions has shifted beyond three qubits. Can our concurrence fill $F_{123}$ and a generalized geometric approach via Concurrences be extended to more-qubit cases? The answer is not presently known, but we believe the answer is yes.
What evidence exists? In the four-qubit case, an attractive generalization of Concurrence Fill is not area of a triangle, but the volume of the newly identified ``concurrence tetrahedron''. The areas of the four surfaces of a concurrence tetrahedron are given by the four squared one-to-other concurrences, see Fig. \ref{fig:tetrahedron} for illustration. This is well defined for the $W$ class states for which the entanglement monogamy relation yields equality (see \cite{coffman2000}). However, the case for other class states is more complicated since four surfaces with given areas cannot determine the volume of a tetrahedron. More criteria or more options such as another minimization procedure are possibly needed and are being considered.

It is also necessary to be reminded of a missing element in the discussion so far: we have not mentioned entanglement measures for mixed-states. However, we can say that this is known to be, at least conceptually, an elementary step, now that the highly satisfactory pure state tripartite entanglement measure $F_{123}$ has been identified. The process of convex-hull extension is well known \cite{uhlmann1998}. Its difficulty is merely computational. Unfortunately, even if completely systematic, the required computation is easily seen to be very challenging \cite{xie2022}.

%In summary, we introduced the background history and the challenging nature of quantum entanglement, including a detailed reminder of its vector-space   and multi-party quantum entanglement and the determination of its measure. explained a recent success in beginning to takeGiven the widely accepted reputation of entanglement as a counter-intuitive feature of quantum theory, we first carefully describe entanglement itself, and describe its measure in the two-qubit pure-state context. Then we introduce the features that have made it challenging for several decades to extend that measure beyond 2-qubit entanglement, and finish with a current understanding that solves the 3-qubit entanglement challenge. This takes into account the fundamental division of the 3-qubit state space into two completely independent sectors identified with the so-called GHZ and $W$ states. 

\section*{Acknowledgements}
We thank M. A. Alonso and X.-F. Qian for several key initial discussions, and we thank Y.-Y. Zhao for valuable conversations.

\section*{Disclosure statement}
The authors report there are no competing interests to declare.

\section*{Funding}
Financial support was provided by National Science Foundation Grants PHY-1501589 and PHY-1539859 (INSPIRE).

\section*{Notes on contributors}

\textbf{\textit{Songbo Xie}} is a PhD Candidate in Physics at the University of Rochester, studying quantum entanglement. He received his B.S. in Physics in 2017 from Shanghai Jiao Tong University in China, supervised by Prof. Zhiguo Lv. His researches cover the geometric quantification of entanglement in multi-qubit systems, bizarre behavior of entanglement in quantum dynamical systems, and the evaluation of entanglement with experimental measurements. \\

\noindent \textbf{\textit{Joseph Eberly}} received his degrees from Penn State and Stanford where E.T. Jaynes was his PhD advisor. He is now the Andrew Carnegie Professor of Physics and Professor of Optics in the University of Rochester where his research in theoretical quantum optics and atomic physics has engaged more than 40 PhD students in interesting topics such as time-dependent spectra, multiphoton ionization, atomic state collapse and revival in cavity QED, first observation of Bessel beams, and sudden death and freezing of quantum entanglement. Professor Eberly is a Fellow of APS and of Optica (formerly OSA) and has served as Chair of the APS Division of Laser Science and President of OSA. He was the Founding Editor of Optics Express. Professor Eberly has received the Distinguished Alumnus Award of the Penn State College of Science, Smoluchowski Medal of the Physical Society of Poland, the Georgen Award for Artistry in Undergraduate Teaching at the University of Rochester, Frederick Ives Medal and Quinn Prize of OSA, and has been elected as Honorary Member of Optica. He is a Foreign Member of the Academy of Science of Poland.

\section*{ORCID}
{\it Songbo Xie} \orcidlink{https://orcid.org} https://orcid.org/0000-0002-9136-9481

\bibliographystyle{triangle}
\bibliography{triangle}
\end{document}